\documentclass[sigconf,screen=true,bookmarks=false]{acmart}

\usepackage[utf8]{inputenc} %
\usepackage[T1]{fontenc}    %
\usepackage{pifont}

\usepackage{algorithm}
\usepackage{algorithmicx}
\usepackage[noend]{algpseudocode}
\usepackage{graphicx}
\usepackage{threeparttable}
\usepackage{multirow}
\usepackage{amsmath}
\usepackage{bm}
\usepackage{bbm}
\usepackage{amsthm}
\usepackage{enumitem} %
\usepackage[subrefformat=parens,labelformat=parens]{subfig}

\usepackage{wrapfig}

\definecolor{citecolor}{RGB}{34,139,34}
\definecolor{mydarkblue}{rgb}{0,0.08,1}
\definecolor{mydarkgreen}{rgb}{0.02,0.6,0.02}
\definecolor{mydarkred}{rgb}{0.8,0.02,0.02}
\definecolor{mydarkorange}{rgb}{0.40,0.2,0.02}
\definecolor{mypurple}{RGB}{111,0,255}
\definecolor{myred}{rgb}{1.0,0.0,0.0}
\definecolor{mygold}{rgb}{0.75,0.6,0.12}
\definecolor{myblue}{rgb}{0,0.2,0.8}
\definecolor{mydarkgray}{rgb}{0.,0.2,0.2}

\definecolor{lightred}{RGB}{255,235,235}
\definecolor{lightgreen}{RGB}{235,255,235}
\definecolor{lightblue}{RGB}{235,235,255}
\definecolor{lightcyan}{RGB}{235,255,255}
\definecolor{lightmagenta}{RGB}{255,235,255}
\definecolor{lightyellow}{RGB}{255,255,235}

\definecolor{qxkcolor}{RGB}{215,235,255}
\definecolor{softmaxcolor}{RGB}{230,235,255}
\definecolor{probxvcolor}{RGB}{255,255,235}

\definecolor{topkcolor}{RGB}{255,235,235}
\definecolor{zecolor}{RGB}{255,255,235}
\definecolor{dynacolor}{RGB}{235,255,255}

\definecolor{reviewcolor}{RGB}{0,0,200}

\newcommand{\ceil}[1]{\lceil #1 \rceil}

\newcommand{\calD}{\mathcal{D}}

\newcommand{\calL}{\mathcal{L}}

\newcommand{\calI}{\mathcal{I}}

\DeclareMathOperator*{\argmax}{argmax}

\theoremstyle{plain}

\theoremstyle{definition}

\newcommand{\squishlist}{
 \begin{list}{$\bullet$}
  { \setlength{\itemsep}{0pt}
     \setlength{\parsep}{3pt}
     \setlength{\topsep}{3pt}
     \setlength{\partopsep}{0pt}
     \setlength{\leftmargin}{1.5em}
     \setlength{\labelwidth}{1em}
     \setlength{\labelsep}{0.5em} } }
     
\newcommand{\squishend}{
  \end{list}  }

\graphicspath{{./figs/}}
\usepackage{geometry}
\copyrightyear{2022}
\acmYear{2022}
\setcopyright{acmcopyright}\acmConference[ASP-DAC '25]{Proceedings of the 30th Asia and South Pacific Design Automation Conference}{Jan. 20--23, 2025}{ Tokyo Odaiba Miraikan, Japan}
\acmBooktitle{Proceedings of the 59th ACM/IEEE Design Automation Conference (DAC) (DAC '22), July 10--14, 2022, San Francisco, CA, USA}
\acmPrice{15.00}
\acmDOI{10.1145/xxxxxxx.xxxxxxx}
\acmISBN{978-1-4503-9142-9/22/07}

\begin{document}
\settopmatter{printacmref=false} %

\pagestyle{plain} %

\title{
The Unlikely Hero: Nonideality in Analog Photonic Neural Networks as Built-in Defender Against Adversarial Attacks
}

\author
{
Haotian Lu,
Ziang Yin,
Partho Bhoumik,
Sanmitra Banerjee,
Krishnendu Chakrabarty,
Jiaqi Gu\\
Arizona State University\\
\small\textit{jiaqigu@asu.edu}
}
\begin{abstract}
\label{abstract}
Electronic-photonic computing systems have emerged as a promising platform for accelerating deep neural network (DNN) workloads. 
Major efforts have been focused on countering hardware non-idealities and boosting efficiency with various hardware/algorithm co-design methods.
However, the adversarial robustness of such photonic analog mixed-signal AI hardware remains unexplored.
Though the hardware variations can be mitigated with robustness-driven optimization methods, malicious attacks on the hardware show distinct behaviors from noises, which requires a customized protection method tailored to optical analog hardware.
In this work, we rethink the role of conventionally undesired non-idealities in photonic analog accelerators and claim their surprising effects on defending against adversarial weight attacks.
Inspired by the protection effects from DNN quantization and pruning, we propose a synergistic defense framework tailored for optical analog hardware that proactively protects sensitive weights via pre-attack unary weight encoding and post-attack vulnerability-aware weight locking.
Efficiency-reliability trade-offs are formulated as constrained optimization problems and efficiently solved offline without model re-training costs.
Extensive evaluation of various DNN benchmarks with a multi-core photonic accelerator shows that our framework maintains \emph{near-ideal} on-chip inference accuracy under adversarial bit-flip attacks with merely $<$3\% memory overhead. 
Our codes are open-sourced at \href{https://github.com/ScopeX-ASU/Unlikely_Hero}{link}.

\end{abstract}

\maketitle

\section{Introduction}
\label{sec:Introduction}

In recent years, analog optical neural networks (ONNs) stand out for their ability to deliver unparalleled speed and efficiency, presenting a promising avenue for artificial intelligence (AI) applications~\cite{NP_NATURE2017_Shen, NP_PIEEE2020_Cheng, NP_NaturePhotonics2021_Shastri, NP_ACS2022_Feng, NP_Science2024_Xu, NP_SciRep2017_Tait, NP_Nature2021_Xu, NP_Nature2021_Feldmann, NP_NatureComm2022_Zhu}.
However, deploying photonic accelerators is impeded by various non-idealities, e.g., low-precision control, hardware noises, and crosstalk, that increase the design complexity to ensure robust deployment.
Extensive prior work has focused on suppressing the physical non-ideality and improving the system robustness via cross-layer hardware/algorithm co-design~\cite{NP_DATE2020_Gu, NP_ICCAD2019_Zhao, NP_ICCAD2020_Zhu,NP_TCAD2022_Mirza}.
Besides the built-in variations/noises, photonic accelerators are exposed to adversarial attacks~\cite{BFADefense_ICCAD2017_Liu,BFA_ICCV2019_Rakin,TBFA_TPAMI22_Rakin}
in real-world deployment, raising hardware security concerns.
Like digital AI accelerators, we envision that malicious attacks, e.g., bit-flip attacks in stored NN weights, will quickly become another potential roadblock for emerging optical analog neural accelerators.
Only tens of bit-flips on the most significant bits (MSB) of critical weights severely degrade the accuracy.
Effective pre-attack protection and post-attack accuracy recovery schemes that leverage the unique properties of analog optical hardware remain unexplored.

Prior work in neural network defense has explored various training-based and training-free defense methods~\cite{NP_DATE2020_Gu, DefendBFA_CVPR2020_He, BFADefense_DAC2020_Li, RADAR_DATE21_Li}.
For example, noise-aware training (NAT)~\cite{NP_DATE2020_Gu} and adversarial training have been proposed to smooth the NN loss landscape and increase attack tolerance.
Among various defense methods, a class that exploits model compression techniques is particularly interesting in the analog NN context.
Quantization, a common model compression method, has been applied for defense. 
Binarization-aware training (BAT)~\cite{DefendBFA_CVPR2020_He}, as a training-based method, has been proposed to provide pre-attack protection by reducing weight sensitivity via 1-bit weights.
However, training-based methods usually suffer from huge model re-training costs and encounter practical concerns in data access, privacy, etc.
As a pre-attack protection conducted offline, training-based defense maximizes the average performance across arbitrary attacks, which usually lack precise protection at the cost of task performance degradation.
Training-free methods usually occur post-attack as a complementary protection mechanism, detecting/localizing the victim weights~\cite{BFADetection_ICCAD2020_Liu} and resuming accuracy by error mitigation/correction.
A representative training-free defense method is pruning-based accuracy recovery~\cite{RADAR_DATE21_Li}.
It detects victim weight groups via MSB checksum verification and prunes detected weights to 0 to partially reduce the bit-flip induced error.
Since low-bit precision and sparsity naturally exist as built-in primitives in optical AI hardware mainly for efficiency-accuracy trade-offs, it inspires us to explore their novel usages in defense.

ONNs' non-idealities have been treated as undesired hardware restrictions compared to digital computers, while in this work, we revisit their role as intrinsic low-cost defenders, adding reliability as a new dimension in the hardware/software co-design space.
In this work, \emph{for the first time}, we propose a synergistic defense framework for photonic AI hardware that provides pre-attack protection via an optics-inspired unary weight representation and post-attack accuracy recovery via a sensitivity-aware on-chip weight locking technique.
Memory efficiency and adversarial robustness are co-optimized to provide near-ideal accuracy protection at marginal memory overhead.

The major contributions of this paper are as follows:

\squishlist
    {\item We investigate the adversarial robustness of optical analog neural networks under malicious weight attacks and explore the built-in protection of the photonic accelerator non-idealities.
    }
    {\item We propose a quantization-inspired \emph{truncated complementary unary weight encoding} to minimize the ONN weight sensitivity with optimized efficiency-robustness trade-offs.
    }
    {\item We propose a pruning-inspired clustering-based \emph{weight locking} technique that co-optimizes detection precision, accuracy recovery, and memory efficiency.
    }
    {\item Our synergistic framework with integrated pre-attack unary protection and post-attack weight locking has shown near-ideal resumed accuracy with a marginal 3\% memory overhead.
    }
\squishend

\section{Preliminaries}
\label{sec:Background}

\subsection{Photonic AI Accelerators and Optical DAC}
\label{sec:ArchSetting}

\begin{figure}
    \centering
    \vspace{-5pt}
    \includegraphics[width=0.9\columnwidth]{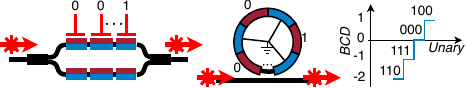}
    \vspace{-10pt}
    \caption{\small (\emph{Left}) Example optical DACs with segmented modulators~\cite{NP_OE2017_Samani,NP_JSSC2017_Moazeni}.
    (\emph{Right}) Signed BCD to unary representation conversion.
    }
    \label{fig:ODAC}
    \vspace{-5pt}
\end{figure}

Various photonic AI accelerators have been demonstrated~\cite{NP_NATURE2017_Shen, NP_PIEEE2020_Cheng, NP_NaturePhotonics2021_Shastri, NP_ACS2022_Feng, NP_DATE2019_Liu,NP_HPCA2024_Zhu}.
As a case study, we focus on one multi-core photonic AI accelerator architecture based on dynamic photonic tensor cores (PTC) \cite{NP_HPCA2024_Zhu}.
Each PTC takes two optically-encoded matrices and performs speed-of-light matrix-matrix multiplication.
The input signals are quantized to reduce the digital-analog conversion (DAC) cost.
The inputs $X$ are quantized to 8-bit fixed-point numbers, while the weights are quantized to $b$-bit, e.g., ranging from 4-bit to 8-bit.
A recent trend to reduce the electrical DAC (eDAC) power bottleneck is to employ optical DAC (oDAC) modules, which encode discretized values to light magnitude with segmented modulators~\cite{NP_OE2017_Samani,NP_JSSC2017_Moazeni}, as shown in Fig.~\ref{fig:ODAC}.

\begin{figure}
    \centering
    \includegraphics[width=\columnwidth]{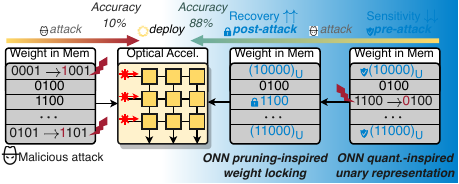}
    \vspace{-20pt}
    \caption{\small Proposed built-in defense flow for photonic AI accelerators against malicious weight attack.}
    \label{fig:teaser}
    \vspace{-10pt}
\end{figure}

The controller in segmented oDAC is partitioned into 2$^{b}-1$ equal-length segments, each contributing to 1 least-significant bit (LSB) of the encoded value.
In this setting, the binary weight value needs to be converted to a unary representation where a '1' applies a voltage to that bit without the need for high-power eDAC. 
Thus, the number of leading 1's can represent the original binary-coded digit (BCD), i.e., 
\begin{equation}
    \small
    (w)_{B}=\{1\}^w\{0\}^{2^b-1-w}=\big({\underbrace{{1,\cdots},1}_{w},\underbrace{0,\cdots,0}_{2^b-1-w}}\big)_{U},
\end{equation}
where $w$ is a signed integer value.
For example, (11110000)$_{U}$=(4)$_{B}$.
Compared to PTCs with high-speed eDACs, oDAC-enhanced designs show significant power reduction.
This \textbf{unique hardware architecture and property inspire us to explore intrinsic unary encoding as an effective protection} mechanism that brings \textbf{minimum weight sensitivity} without extra BCD-to-Unary conversion cost, as unary coding is the \textbf{built-in primitive}.

\section{Proposed Defense Framework}
\label{sec:Method}
We will introduce the threat model and investigate built-in defense mechanisms in non-ideal analog photonic accelerators.
As shown in the overview Fig.~\ref{fig:teaser}, two key techniques will be introduced to provide both pre-attack weight protection and post-attack accuracy recovery with optimized memory-robustness trade-offs.

\subsection{Threat Model and Attacker Settings}
\label{sec:Threatmodel}
As a case study, we assume a widely employed attacker model: \textbf{gradient-based attacker} BFA~\cite{BFA_ICCV2019_Rakin}.
Important assumptions on the threat model are given in Table~\ref{threatmodel}. 
We follow the standard white-box attack threat model assumptions as previous work~\cite{BFA_ICCV2019_Rakin, TBFA_TPAMI22_Rakin, DefendBFA_CVPR2020_He, BFADefense_TC2022_Liu}.
\begin{table}[h]
\centering
    \vspace{-10pt}
    \caption{\small Threat model assumed in this work.}
    \vspace{-10pt}
    \resizebox{0.95\columnwidth}{!}{
    \begin{tabular}{|c|c|}
    \hline
    \textbf{Access Required} & \textbf{Access NOT Requied} \\ \hline
    DNN model and parameters & Training Configurations     \\
    A mini-batch of attack dataset  & Modify scaling factors in quantization \& Norm.   \\
    On-chip forward/backward prop.      & Modify address mapping/look-up tables    \\ \hline
    \end{tabular}
}
\label{threatmodel}
\end{table}

Eq.~\eqref{eq:our_attacker} describes the target of an on-chip adversarial attacker under Hamming Distance (HD) and inference budget ($T_{inf}$) constraint.
\begin{equation}
\small
    \begin{aligned}
    \label{eq:our_attacker}
    \min_{\calI_{A}}~Acc(\widehat{W}_{\calI_A}, \calD^{test})&\approx \max_{\calI_{A}}~\calL(\widehat{W}_{\calI_A}, \calD^{att})\\
    \text{s.t.}~~ \|\widehat{W}_{\calI_A}-W\|_1 \leq HD;& \quad \# \text{ of model inferences}\leq T_{inf}, 
    \end{aligned}
\end{equation}
where $\calI_A$ is the selected bits to attack, $\widehat{W}_{\calI_A}$ is the attacked weights, and $\calD^{att}$ is the attack dataset, usually a small batch of $BS$ examples.
The minimization of post-attack test accuracy is often estimated by maximizing the loss function on a small attack dataset.

\noindent\underline{Gradient-based Attacker (BFA)}.~
The gradient-based attacker has access to the gradient information of a mini-batch of data via on-chip backpropagation and attacks the most sensitive bits.
Specifically, we adopt the attacker algorithm BFA~\cite{BFA_ICCV2019_Rakin} that progressively searches for the most sensitive bits indicated by the largest absolute gradient if the flip direction aligns with the gradient.
The gradients of all weights will be re-evaluated every time it flips one bit. 
Each bit-flip requires forward and backward propagation, equivalently consuming three inference budgets. 
If the attacker consumes all inference budget but still has an extra hamming distance budget left unused, it will directly select the most sensitive but unattacked weights to flip their MSB to make sure it \emph{always uses up the hamming distance budget}. We conduct all the experiments under the $HD=100$ condition.

\subsection{Efficient Built-in Pre-Attack Defense via \textbf{\textit{Unary}} Weight Representation}
For efficiency and control complexity consideration, the weights of photonic analog AI hardware are often quantized to low-bitwidth fixed-point numbers~\cite{NP_DATE2020_Gu, NP_HPCA2024_Zhu}, usually represented as binary-coded decimal (BCD) format with 2's complement encoding.
The weights are fetched from electrical memory, converted to voltage signals via DACs, and encoded in the optical domain for computing.
To investigate the role of quantized weight encoding in the adversarial robustness of analog hardware, we first raise several critical questions: 
\ding{202} \textbf{How does quantization impact the adversarial robustness against bit-flip attack}?
\ding{203} \textbf{How can we leverage the natural unary representation inspired by optical DAC as an effective defense}?
\ding{204} \textbf{What is the robustness-memory trade-off of unary representation and how to avoid the exponential memory cost}?

\begin{figure}
    \centering
    \vspace{-15pt}
    \subfloat[]{\includegraphics[width=0.42\columnwidth]{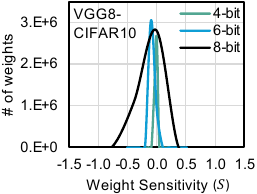}
    \label{fig:BitSensitivity}}
    \subfloat[]{\includegraphics[width=0.49\columnwidth]{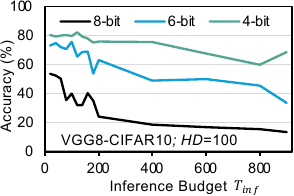}
    \label{fig:QuantizationRobustness}}
    \vspace{-10pt}
    \caption{\small (a) Lower bitwidth reduces weight sensitivity. 
    (b) Low-bit quantization helps improve bit-flip attack robustness.}
    \vspace{-10pt}
\end{figure}

\subsubsection{Protection Effects of Quantization}
To answer the question \ding{202}, we investigate how sensitivity changes with various bitwidth in quantization.
In Fig.~\ref{fig:BitSensitivity}, we observe that low-bit quantization can reduce the overall weight sensitivity defined later in Eq.~\eqref{eq: Sensitivity-Taylor}.
Hence, in Fig.~\ref{fig:QuantizationRobustness}, we observe a clear protection effect from low-bit quantization against bit-flip attack, which lays the foundation for our further study in memory-efficient unary representation.

\subsubsection{Unary Representation as Built-in Protection}
BCD-format is \emph{compact in storage but sensitive to bit-flip attack} since the MSB flip can cause significant deviation by half of the weight range, which casts a serious reliability threat to the hardware.
An intuitive solution is to leverage the built-in \emph{unary representation} to minimize the bit-flip sensitivity as all bits in unary-coded weight are LSB.
With a predefined protection rate $\alpha$, i.e., the percentage of weights protected by unary representation, the protected weights can be searched by maximizing the post-attack accuracy, 
which reflects the protection effectiveness,
\begin{equation}
    \small
    \label{eq:UnaryOptimization}
    \begin{aligned}
      \calI_{U}^* &= \argmax_{\calI_U}~ Acc(\widehat{W}_{\calI_{U}}, \calD^{val}),  
    \end{aligned}
\end{equation}
where $\mathcal{I}_{U}^*$ is the selected indices for unary protection to maximize the validation accuracy after attack, $\widehat{W}_{\mathcal{I}_{U}}$ represents the attacked weights.
We denote the number of weights protected as $N_U=\ceil{\alpha|W|}=|\mathcal{I}_{U}^*|$.

For the original unary representation, the memory overhead is exponential, which limits protection efficiency, as shown in Eq.~\eqref{Eq:TCNOO}.
\begin{equation}
\small
    \label{Eq:TCNOO}
    m_{U} = \big((2^b - 1)|\calI_U| + \sum_{l=1}^L \lceil \log_2 N_U^l\rceil \times |\mathcal{I}_{U,l}|\big)/(b|W|),
\end{equation}

Given the memory overhead budget, we can roughly derive the maximum number of weights we can protect.
Then, the next phase is to \textbf{determine the weights to protect}.
The overall pre-attack protection algorithm with unary representation is detailed in Alg.~\ref{algo:Quasi-Unary}. 
To maximize the protection effectiveness, we prefer to protect vulnerable weights that show the largest bit-flip sensitivity $S$.

\noindent\underline{Bit-flip-Aware Weight Sensitivity Evaluation}.~
A widely used weight sensitivity is the magnitude of the first-order gradient $|\nabla_{W}\calL|$ or second-order gradient $|\nabla^2_{W}\calL|$ in the literature.
Those metrics are designed for small random perturbations in a neighbor region where the gradient and curvature information can capture the sensitivity.
However, the bit-flip attack is not a random perturbation that has \textbf{a determined direction}, i.e., from 0/1 to 1/0; meanwhile, the large deviation from the MSB flip \textbf{breaks the assumption of small local perturbation}.
Therefore, we employ a bit-flip-aware sensitivity score based on Taylor expansion of the loss on the validation dataset,
\begin{equation}
\label{eq: Sensitivity-Taylor}
\small
    S=\calL - \calL_0 \approx  
    \nabla_W\calL \cdot \Delta W_{MSB}
    + \frac{1}{2}\cdot \nabla_W ^2\calL \cdot \Delta W_{MSB}^2,
\end{equation}
where the Hessian matrix is approximated by its diagonal entries $\nabla_W ^2\calL$, and $\Delta W_{MSB}$ is the perturbation caused by MSB-1 flip.
A larger sensitivity $S$ represents a higher vulnerability to bit-flip.
This score is aware of the alignment of the bit-flip direction with the gradients.
Only bit-flips leading to larger S will be considered for protection.

\noindent\underline{Sensitivity-Guided Memory Overhead Assignment}.~
Once we obtain the sensitivity scores for all weights, we need to further determine how to leverage the scores as guidance to distribute the memory overhead budget to all neural network layers.
We propose \textbf{Top-Sensitive-Layer Assignment} that ranks layers based on their overall sensitivity and allocates all memory budgets to the most sensitive layers.
As shown in Fig.~\ref{fig:LayerSensitivity}, layer sensitivity is estimated by the averaged 50\%-quantile and 75\%-quantile of sensitivity of all weights, i.e., $\bar{S}^l=(Q_{50\%}(S)+Q_{75\%}(S))/2$.
The sorted layer indices from most sensitive to least sensitive are $(l_1,l_2,\cdots,l_L)$.
Formally, the layer-wise overhead budgets are $(\frac{|W^{l_1}|}{\sum_{i=1}^L|W^{l_{i}}|}, \frac{|W^{l_2}|}{\sum_{i=1}^L|W^{l_{i}}|}, \cdots, m-\sum_{j=1}^{L'}\frac{|W^{l_j}|}{\sum_{i=1}^L|W^{l_{i}}|},0,\cdots,0)$, where only the top-$L'$ sensitive layers can have weight protection.

\noindent\underline{Attack-injected Search for Weight Selection}.~
Based on the sensitivity-weighted guidance, we can select a certain number of vulnerable weights to protect for each layer.
However, solely relying on weight sensitivity ranking to select weights to protect does not directly optimize toward the true objective, i.e., maximization of post-attack accuracy $Acc(\widehat{W})$.
Since the selection procedure is pre-deployment (offline), we can afford to search by sampling weights with sensitivity-weighted probability and select the group that brings the best protection based on bit-flip attack emulation and validation accuracy.

\begin{algorithm}
    \caption{\small Pre-attack unary weight protection algorithm}
    \label{algo:Quasi-Unary}
    \begin{algorithmic}
    \small
    \Require Loss function $\calL(W)$, protection rate $\alpha$, hamming distance for attacker $HD$, \# of attacks $T_a$, max search steps $T$, and validation set $\calD^{val}$.
    \State \text{Calculate per-weight sensitivity} $\{S^l\}_{l=1}^L$ and layer sensitivity $\{\bar{S}^l\}_{l=1}^L$
    \State $\{N_{U}^{l}\}_{l=1}^{L'}\gets \texttt{mem\_assignment}( \alpha, \{\bar{S}^l\}_{l=1}^L)$
    \For{$l \gets 1\cdots L'$}
    \State Best accuracy $Acc^* \gets 0$
    \For{$t\gets 1\cdots T$}
    \State Sample $N_{U}^{l}$ indices with probability $P^l=\texttt{softmax}(S^l)$ as $\calI_{U, t}^{l}$
    \State Protect weights with TCU: $W_{\calI_{U,t}^l}^l \gets \texttt{BCD-to-TCU}(W^l, \calI_{U,t}^l)$
    \For{$j \gets 1 \cdots T_a$}
    \State $\widehat{W}_{\calI_{U,t}^l}^l\gets \texttt{Attack}(W_{\calI_{U,t}^l}^l, HD);\quad Acc_j \gets Acc(\widehat{W}^l_{I_{U, t}^{l}},\calD^{val})$
    \EndFor
    \State Get worst post-attack accuracy: $Acc_{t}\gets \min \{Acc_j|\forall j \in [T_a]\}$
    \If{$Acc_t > Acc^*$}
    \State Record most protective weights: $Acc^* \gets Acc_t; ~~ \calI_{U}^{l*} \gets \calI_{U,t}^l$
    \EndIf
    \EndFor
    \EndFor
    \Ensure $\calI_{U}^*\gets \{\calI_{U}^{l*}|l\in[L']\}$
    \end{algorithmic}
\end{algorithm}

\subsubsection{Memory-Efficient Truncated Complementary Unary Representation}
\begin{figure}
    \centering
    \includegraphics[width=\columnwidth]{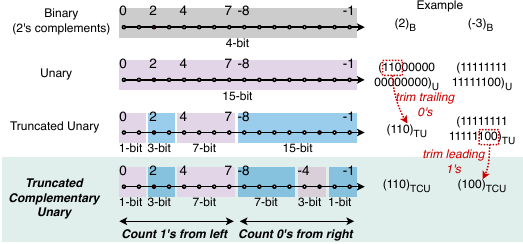}
    \vspace{-15pt}
    \caption{\small Different coding formats. Our truncated complementary unary representation shows superior memory efficiency.}
    \label{fig:TCU}
    \vspace{-10pt}
\end{figure}

To answer question \ding{204}, we propose a memory-efficient truncated complementary unary (TCU) representation. 
An important observation is that in the original unary representation, the \textbf{large number of trailing zeros for small values are only for bitwidth alignment purposes without any expressiveness}.
Hence, an intuitive compression method is to truncate the trailing zeros to reduce bitwidth from $(2^b-1)$ to $\hat{b}$.
For example, if $b=3$, we can compress numbers smaller than 4 by using only 4-bit in truncated unary-format instead of $2^3-1=7$-bit, e.g., $(7b'1100000)_U\xrightarrow{truncate} (4b'1100)_{U}$, without changing its actual encoded value of $(2)_B$.

\noindent\underline{How to select optimal truncation bitwidth $\hat{b}$?} -- If $\hat{b}$ is too large, not many zeros can be truncated.
On the other hand, if $\hat{b}$ is overly small, only a few weights with small values can be expressed by the truncated bitwidth.
Both cases give unsatisfactory memory saving.

We first illustrate a truncated version of unary format (TU) by clustering weights into exponentially-spaced bins and assigning a truncation bitwidth to cover the largest value in each bin, as shown in Fig.~\ref{fig:TCU}.
For small positive values, such a method can significantly trim the redundant trailing zeros for memory reduction.
But it is not very efficient for the largest bin.
Moreover, since most sensitive weights have small absolute values~\cite{RADAR_DATE21_Li}, a large proportion of negative weights, unfortunately, fall into the largest bin.

\begin{figure}
    \centering
    \vspace{-15pt}
    \subfloat[]{\includegraphics[width=0.467\columnwidth]{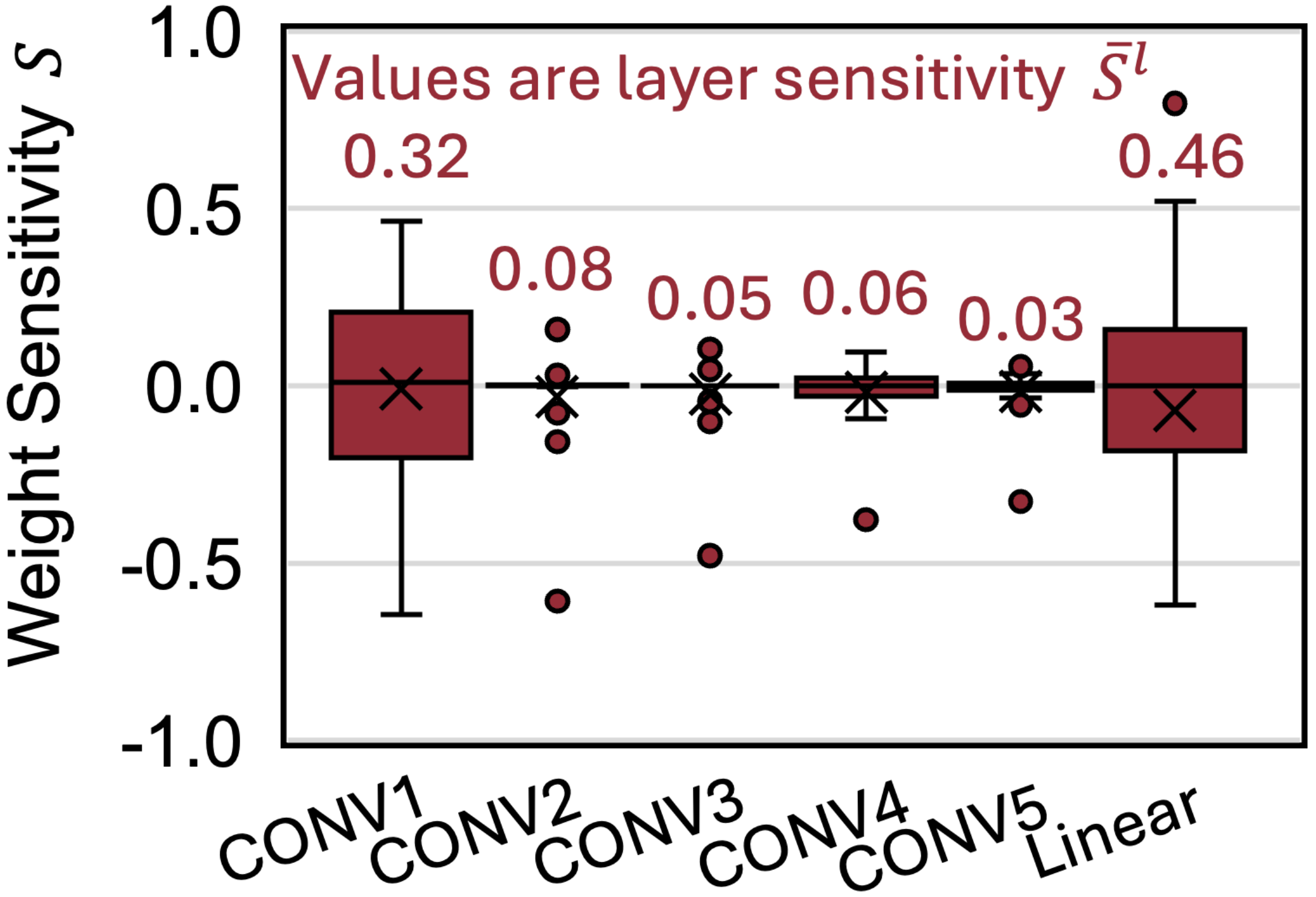}
    \label{fig:LayerSensitivity}}
    \subfloat[]{\includegraphics[width=0.53\columnwidth]{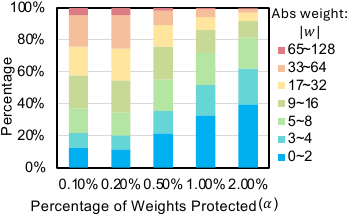}
    \label{fig:WeightDistribution}}
    \vspace{-10pt}
    \caption{\small (a) 6 layers in 8-bit VGG-8 shows distinct layer sensitivity statistics.
    (b) 
    Distribution of absolute values of weights protected by Unary Protection for 8-bit VGG-8 on CIFAR10.
    Vulnerable weights that deserve to be protected have small magnitudes.}
    \vspace{-10pt}
\end{figure}

Aware of the \textbf{Gaussian-like weight distribution} in real neural networks 
and the important property of unary representation, i.e., \textbf{counting 0's is equivalent to counting 1's}, we propose a complementary unary format (TCU) that stores trailing 0's and trims leading 1's for negative values.
For instance, the required bitwidth for -3 can be reduced from 15-bit in TU-format to 3-bit in TCU-format, as illustrated in Fig.~\ref{fig:TCU}.
Similar to logarithmic quantization, the exponentially-sized bins in TCU-format 
reduce the bin count while minimizing memory overhead by holding a large number of \textbf{small-value yet sensitive weights in the lowest-bitwidth bins}, shown in Fig.~\ref{fig:WeightDistribution}.

The memory overhead ratio $m_{TCU}$ of TCU-format and indexing overhead is formulated in Eq.\eqref{Eq:TCU}.
\begin{equation}
\small
    \label{Eq:TCU}
    m_{TCU} = \sum_{l}^{L'}\Big(\sum\limits_{i\in I_U^l} 2^{\lceil \log_2 \min(2 ^ b - |W_i|, |W_i|)\rceil} + \lceil \log_2 N_l\rceil \times |\mathcal{I}_U^l|\Big)/(b\sum_{l}^{L}|W^l|).
\end{equation}

\vspace{-10pt}
\subsection{Post-attack Accuracy Recovery via Sensitivity-aware Weight Locking}
Pre-attack unary protection is unaware of the actual attacked bits as it is performed offline before deployment.
\textbf{This lack of precise targets makes the coverage of pre-attack protection insufficient} if only a small percentage ($\alpha$) of weights can be converted to TCU format.
It is necessary to employ post-attack detection and recovery mechanisms to compensate for this inevitable protection miss.

Pruning is widely used in analog ONNs to improve energy efficiency~\cite{NP_TCAD2020_Gu,NP_JSTPE2023_Banerjee,NP_ICCAD2024_Yin,NP_isvlsi2022_banerjee2022pruning,NP_OFC2022_banerjee2022champ}.
We ask two critical questions: \ding{202} \textbf{how to leverage the natural hardware sparsity for defense?} and \ding{203} \textbf{how to trade off pruning-induced accuracy loss and protection effects?} 

In previous work~\cite{RADAR_DATE21_Li}, pruning is utilized to force detected under-attack weight groups to zero to partially cancel out the bit-flip errors.
This method is intuitive due to two facts.
(1) First, the MSB flip creates a deviation of half of the weight range, which always changes the sign of the weight, e.g., $(-3)_B \rightarrow (5)_B$ for a 4-bit weight.
Hence, forcing it to 0 always reduces the error $|(-3)-0|<|(-3)-5|$, at least on the weight itself.
(2) Second, weight distribution shown in Fig.~\ref{fig:WeightDistribution} shows that many sensitive weights under attack have small magnitudes~\cite{RADAR_DATE21_Li}, which further justifies that pruning is a promising built-in mechanism for accuracy recovery, which answers question~\ding{202}.

However, simple weight pruning fails to resume accuracy in practice because many of the pruned weights are either \textbf{still far away from 0} or not real victim weights due to inevitable \textbf{false alarms in group-wise detection}.
In other words, pruning fake victims to 0 turns out to be a self-attack.

To answer question \ding{203}, we propose sensitivity-aware weight locking, which \emph{generalizes prior pruning-based method} and significantly \emph{boosts the protection effectiveness} with optimal clustering and locking.

\begin{figure}
    \centering
    \vspace{-5pt}
    \includegraphics[width=\linewidth]{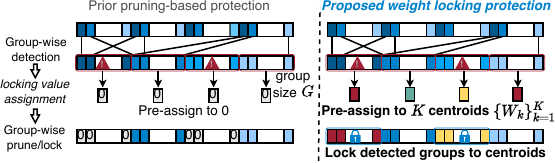}
    \vspace{-20pt}
    \caption{\small Comparison between pruning-based protection and our proposed weight-locking method.}
    \label{fig:LockingIllustration}
    \vspace{-10pt}
\end{figure}

\begin{figure}
    \centering
    \includegraphics[width=0.85\columnwidth]{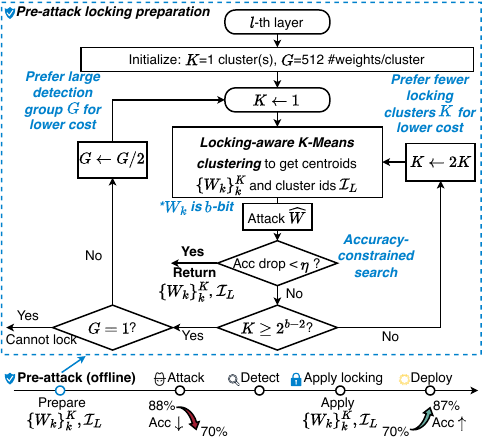}
    \vspace{-10pt}
    \caption{\small The layer-wise offline search procedure to find optimal detection group size and locking solutions given accuracy constraints.}
    \label{fig:WeightLocking}
    \vspace{-5pt}
\end{figure}

Post-attack accuracy recovery generally follows two steps: detection (localization) and resume.
We assume the same detection technique based on group-wise MSB checksum verification~\cite{RADAR_DATE21_Li}.
A mismatch in checksum will mark the entire group of size $G$ as the victim weight group.
All weights in victim groups will be resumed in the second step.
Shown in Fig.~\ref{fig:LockingIllustration}, unlike the prior pruning method that forces victim weights to 0, we propose sensitivity-aware weight locking that intelligently finds $K$ centroids before deployment and locks detected victim weights to their centroids to maximize recovery effectiveness.
Key trade-offs here include detection group size $G$, which impacts detection accuracy and memory, and cluster number $K$, which impacts both the centroid storage cost and resumed accuracy.

For the $l$-th layer, we formulate it as an accuracy-constrained memory overhead minimization problem as follows in Eq.\eqref{eq:lock-optimization},
\begin{equation}
\small
    \begin{aligned}
        \min_{\{W_k\}_{k=1}^K,\calI_L, G, K} & m_L(G, K)=
        \begin{cases}
            &\frac{|W|(\log_2K+2)}{G\cdot b|W|},~~G > 1\\
            &\frac{|W|(\log_2K + 1)}{b|W|},~~G = 1\\
        \end{cases}\\
        \text{s.t.}~~&Acc_0 - Acc(\widehat{W}_{\calI_L,\{W_k\}_{k=1}^K}) < \eta
    \end{aligned},
    \label{eq:lock-optimization}
\end{equation}
where $W_{k}$ is the $b$-bit centroid for the $k$-th cluster, $\calI_L\in\{1,2,\cdots,K\}^{|W|}$ is the assigned cluster IDs for weights, and $\eta$ is the threshold of the gap between ideal and resumed accuracy.
The above optimization is performed independently for each layer.
For the memory overhead $m_L$: 1) $(\log_2K)/G$ denotes the number of bits required to store the cluster ID for each weight.
2) $1$ or $2/G$ denotes the memory required to store the golden signature used in checksum-based detection~\cite{RADAR_DATE21_Li}. 

The algorithm to solve this optimization problem for layer-$l$ is illustrated in Fig.~\ref{fig:WeightLocking}.
To prioritize the lowest-memory solution, we initialize it to the largest group size $G$=512 and minimum cluster count $K$=1, and gradually search feasible solutions by 
halving $G$ and double increasing $K$.
To find the cluster centroids that \emph{minimize locking-induced accuracy loss}, we augment conventional K-Means clustering to a locking-aware variant.
We first perform single-cluster ($K$=1) K-Means within each group.
The distance $d_{in}$ from weight $W_i$ to centroid $\widetilde{W_n}$ of $n$-th detection group is redefined as in Eq.\eqref{eq:Distance} 
\begin{equation}
    \small
    \label{eq:Distance}
    d_{in}=\nabla_{W_i}\calL \cdot (W_i-\widetilde{W_n})
    + \frac{1}{2}\cdot \nabla_{W_i} ^2\calL \cdot (W_i-\widetilde{W_n})^2.
\end{equation}
We then obtain $N=\ceil{|W|/G}$ group centroids aware of locking errors $Acc(W)-Acc(\widetilde{W})$.
Since $N \gg K$, we further perform a standard K-Means clustering to the obtained $N$ centroids and get $\{W_k\}_{k=1}^K$.

\vspace{-3pt}
\subsection{Synergistic Protection with Integrated TCU Encoding and Weight Locking} 
To provide double protection against bit-flip attacks, we leverage both pre-attack unary protection and post-attack weight locking with co-optimized memory overhead.  
Before deployment, given a protection rate $\alpha$, we first perform pre-attack unary protection in Alg.~\ref{algo:Quasi-Unary} to find the weight indices $\calI_U$ and protect them by converting them to TCU-format, effectively reducing the sensitivity of 
vulnerable weights from MSB to LSB.
Meanwhile, we prepare the locking solutions $(\{W_k\}_{k=1}^K, \calI_L)$ for each layer using the procedure in Fig.~\ref{fig:WeightLocking} given a target accuracy drop threshold $\eta$.
After the attack happens, checksum-based detection is applied to pinpoint potential victim weight groups under adversarial attack.
Then, all weights in the detected groups will be locked to their pre-assigned centroid for post-attack accuracy recovery.
A carefully selected $(\alpha, \eta)$ setting gives the best post-attack accuracy and lowest memory overhead.
Since the memory overhead tends to become very large in two extreme cases, i.e., pure unary protection or pure locking, the optimal solution in the middle range can be simply found by greedy search.
We gradually reduce the unary protection rate $\alpha$, e.g., from 2\% to 0.25\%, and for each $\alpha$, we evaluate the overall memory cost and resumed accuracy for all $\eta$ candidates, e.g., $\eta\in\{1\%,1.5\%,2\%\}$.
The search is stopped when the memory overhead increases, and the most efficient solution can be selected.

\vspace{-5pt}
\section{Experimental Results}
\label{sec:ExperimentalResults}
\subsection{Experiment Setup}
\label{sec:ExpSetup}
\begin{table}
\centering
    \caption{\small Impact of batch-size $BS$ (size of $\calD^{att}$) on post-attack accuracy (8-bit VGG8-CIFAR10) with different inference budgets $T_{inf}$. Accuracy with the same color corresponds to the same hardware cost ($BS\times T_{inf}$). Bold texts show the lowest accuracy with the same color.
    }
    \vspace{-10pt}
    \resizebox{0.8\linewidth}{!}{
        \begin{tabular}{cccccccc}
        \hline
            \multicolumn{1}{c|}{} &
              \multicolumn{7}{c}{Inference Budget $T_{inf}$} \\
            \multicolumn{1}{c|}{\multirow{-2}{*}{Batch Size}} &
              20  &
              40  &
              80  &
              160  &
              320 &
              640  &
              1280 \\ \hline
            \multicolumn{1}{c|}{8} &
              61.28 &
              60.47 &
              \cellcolor[HTML]{F898A2}\textbf{38.15} &
              \cellcolor[HTML]{F1BDA6}38.81 &
              \cellcolor[HTML]{F5DBA4}\textbf{39.39} &
              \cellcolor[HTML]{FFFAB9}32.84 &
              \cellcolor[HTML]{C0F1C8}35.75 \\
            \multicolumn{1}{c|}{16} &
              53.62 &
              \cellcolor[HTML]{F898A2}52.45 &
              \cellcolor[HTML]{F1BDA6}\textbf{35.64} &
              \cellcolor[HTML]{F5DBA4}40.30 &
              \cellcolor[HTML]{FFFAB9}\textbf{20.49} &
              \cellcolor[HTML]{C0F1C8}\textbf{14.68} &
              \cellcolor[HTML]{94DCF8}16.72 \\
            \multicolumn{1}{c|}{32} &
              \cellcolor[HTML]{F898A2}62.91 &
              \cellcolor[HTML]{F1BDA6}50.76 &
              \cellcolor[HTML]{F5DBA4}46.01 &
              \cellcolor[HTML]{FFFAB9}28.12 &
              \cellcolor[HTML]{C0F1C8}17.50 &
              \cellcolor[HTML]{94DCF8}\textbf{15.28} &
              \cellcolor[HTML]{A6C9EC}\textbf{13.06} \\
            \multicolumn{1}{c|}{64} &
              \cellcolor[HTML]{F1BDA6}64.43 &
              \cellcolor[HTML]{F5DBA4}49.26 &
              \cellcolor[HTML]{FFFAB9}36.24 &
              \cellcolor[HTML]{C0F1C8}22.27 &
              \cellcolor[HTML]{94DCF8}16.84 &
              \cellcolor[HTML]{A6C9EC}13.39 &
              28.04 \\
            \multicolumn{1}{c|}{128} &
              \cellcolor[HTML]{F5DBA4}65.35 &
              \cellcolor[HTML]{FFFAB9}50.16 &
              \cellcolor[HTML]{C0F1C8}42.74 &
              \cellcolor[HTML]{94DCF8}26.87 &
              \cellcolor[HTML]{A6C9EC}17.49 &
              \textbf{12.42} &
              11.08 \\
            \hline
        \end{tabular}
        }
    \vspace{-12pt}
    \label{tab:minibatch}   
\end{table}

\begin{table}
\caption{\small Defense efficiency of unary protection with two different methods across different protection rate $\alpha$ and attacker inference budget $T_{inf}$. 
Accuracy is for 8-bit VGG8-CIFAR10. $BS$=16 for BFA attacker.
}
\resizebox{0.95\linewidth}{!}{
\begin{tabular}{c|c|c|c|c}
\hline
\multirow{2}{*}{Method}                                                                    & \multirow{2}{*}{\begin{tabular}[c]{@{}c@{}}Memory \\ Overhead $m_{U}$\end{tabular}} & \multirow{2}{*}{\begin{tabular}[c]{@{}c@{}}Protected Weight \\ Percentage $\alpha$\end{tabular}} & \multicolumn{1}{l|}{\multirow{2}{*}{Worst Acc.}} & \multirow{2}{*}{Mean Acc.} \\
                                                                                           &                                                                                     &                                                                                                  & \multicolumn{1}{l|}{}                            &                            \\ \hline
\multirow{4}{*}{\begin{tabular}[c]{@{}c@{}}Even \\ Assignment\end{tabular}}                & 3.34\%                                                                              & 0.10\%                                                                                           & 19.65                                            & 39.21                      \\
                                                                                           & 8.51\%                                                                              & 0.25\%                                                                                           & 53.48                                            & 69.69                      \\
                                                                                           & 33.90\%                                                                             & 1.00\%                                                                                           & 79.59                                            & 83.12                      \\
                                                                                           & \textbf{309.08\%}                                                                   & \textbf{9.00\%}                                                                                  & 87.16                                            & \textbf{87.31}             \\ \hline
\multirow{4}{*}{\begin{tabular}[c]{@{}c@{}}Top-Sensitive-Layer \\ Assignment\end{tabular}} & 3.34\%                                                                              & 0.10\%                                                                                           & 75.43                                            & 80.06                      \\
                                                                                           & 6.70\%                                                                              & 0.20\%                                                                                           & 78.86                                            & 83.27                      \\
                                                                                           & 16.96\%                                                                             & 0.50\%                                                                                           & 80.02                                            & 84.03                      \\
                                                                                           & \textbf{67.90\%}                                                                    & \textbf{2.00\%}                                                                                  & 87.14                                            & \textbf{87.27}             \\ \hline
\end{tabular}
}
\label{tab:QU-Even}
\vspace{-10pt}
\end{table}

\noindent\textbf{Dataset and NN Models.}~ 
We evaluate our method on VGG-8 CIFAR-10 and ResNet-18 CIFAR-100 for image classification. 
We choose 4-bit, 6-bit, and 8-bit for weight quantization. 
Input activations are 8-bit. 

\noindent\textbf{Training Settings.}~
We pre-train all models for 200 epochs with an Adam optimizer with a 2E-3 learning rate, a cosine decay scheduler, 1E-4 weight decay, and data augmentation (random crop and flip). BatchNorm layers are all frozen after pretraining.

\noindent\textbf{Benchmarks and Metrics}.~
We adopt the strongest attacker setting shown in Section~\ref{sec:AblationAttacker} with $HD$=100.
To cover different attack budget scenarios, we sweep over budgets $T_{inf}$ such that the number of bits flipped ranges from 2 to 
100.
We show averaged/worst/best inference accuracy on 5 attack datasets across all budgets.
Gaussian weight noises (std.=0.005) are injected into all on-chip computations for attackers.
We compare ours to two types of defense baselines.

\emph{(1) Training-based Defense} -- 
Binarization-aware Training (BAT)~\cite{DefendBFA_CVPR2020_He} quantized all weights to 1-bit to reduce bit-flip sensitivity. 
Also, we compare ours to noise-aware training (NAT)~\cite{NP_DATE2020_Gu}, which is usually used to boost ONN robustness.

\emph{(2) Other Training-free Defense} -- We use a previous pruning-based protection method~\cite{RADAR_DATE21_Li} as another training-free baseline representing a special case in our weight locking, i.e., centroids are fixed to 0.

\vspace{-3pt}
\subsection{Ablation Study}
\subsubsection{Batch Size $BS$ and Inference Budget $T_{inf}$}
\label{sec:AblationAttacker}
To make sure we evaluate our method against the \emph{strongest} attacker model, we first evaluate the most efficient batch size settings across different inference budgets in Table~\ref{tab:minibatch}.
We can conclude that 16 images are enough for the attacker to get informative sensitivity scores via stochastic gradient calculation to perform an effective bit-flip attack, which leads to the lowest post-attack accuracy given the same hardware cost ($BS\times T_{inf}$).
An overly small $BS$ gives inaccurate gradients, while too many images consume the inference budget rapidly, helping the attacker marginally.

\begin{table}
\centering
\small
    \caption{\small Memory overhead required by TCU-format and unary representation on 8-bit VGG8-CIFAR10.}
    \vspace{-10pt}
        \resizebox{\linewidth}{!}{
        \begin{tabular}{c|cccccc}
        \hline
        \multicolumn{1}{l|}{Protected Weight Percent $\alpha$} & 0.05\%           & 0.10\%           & 0.20\%           & 1.00\%           & 2.00\%           & 4.00\%            \\ \hline
        Unary Encoding: $m_{U}$ & 1.66\%   & 3.34\%   & 6.70\%   & 33.90\%  & 67.90\% & 136.39\% \\
        Proposed TCU: $m_{TCU}$                                  & \textbf{0.17\%} & \textbf{0.51\%} & \textbf{1.07\%} & \textbf{3.39\%} & \textbf{6.03\%} & \textbf{12.70\%} \\\hline
        Reduction ($m_{U}/m_{TCU}$)               & 9.86$\times \downarrow$ & 6.51$\times \downarrow$ & 6.24$\times \downarrow$ & 9.99$\times \downarrow$ & 11.26$\times \downarrow$ & 10.74$\times \downarrow$ \\ \hline
        \end{tabular}
        }
    \label{tab:CQU-res}
    \vspace{-10pt}
\end{table}

\begin{table}
\centering
\small
    \caption{\small Results of post-attack accuracy recovery by proposed weight locking. 8-bit VGG-8 on CIFAR-10 is evaluated. $G$ and $K$ show solutions for all 6 convolutional and linear layers.}
    \vspace{-10pt}
    \resizebox{\linewidth}{!}{
        \begin{tabular}{c|c|c|ccccc|c}
        \hline
        \multirow{2}{*}{$\eta$ (\%)} &
          \multirow{2}{*}{\begin{tabular}[c]{@{}c@{}}Layer-wise Weight \\ Locking Solutions\end{tabular}} &
          \multirow{2}{*}{\begin{tabular}[c]{@{}c@{}}Mem. OV \\ $m_L$\end{tabular}} &
          \multicolumn{5}{c|}{Inference Budget $T_{inf}$} &
          \multirow{2}{*}{\begin{tabular}[c]{@{}c@{}}Mean \\ Acc.\end{tabular}} \\
         &
           &
           &
          20 &
          80 &
          160 &
          400 &
          900 &
           \\ \hline
            - &
          w/o locking &
          0.00\% &
          53.62 &
          35.64 &
          40.30 &
          18.59 &
          13.52 &
          32.33 \\ \hline
        \multirow{2}{*}{\textbf{1}} &
          \textbf{G={[}1, 1, 4, 16, 128, 2{]}} &
          \multirow{2}{*}{\textbf{1.29\%}} &
          \multirow{2}{*}{\textbf{83.49}} &
          \multirow{2}{*}{\textbf{80.77}} &
          \multirow{2}{*}{\textbf{78.36}} &
          \multirow{2}{*}{\textbf{84.60}} &
          \multirow{2}{*}{\textbf{86.59}} &
          \multirow{2}{*}{\textbf{82.76}} \\
         &
          \textbf{K={[}8, 2, 1, 1, 2, 16{]}} &
           &
           &
           &
           &
           &
           &
           \\ \hline
        \multirow{2}{*}{1.5} &
          G={[}1, 1, 8, 16, 128, 2{]} &
          \multirow{2}{*}{1.15\%} &
          \multirow{2}{*}{79.82} &
          \multirow{2}{*}{80.96} &
          \multirow{2}{*}{76.67} &
          \multirow{2}{*}{72.01} &
          \multirow{2}{*}{86.59} &
          \multirow{2}{*}{79.21} \\
         &
          K={[}4, 1, 1, 1, 2, 1{]} &
           &
           &
           &
           &
           &
           &
           \\ \hline
        \multirow{2}{*}{2} &
          G={[}1, 2, 16, 32, 128, 2{]} &
          \multirow{2}{*}{0.97\%} &
          \multirow{2}{*}{78.84} &
          \multirow{2}{*}{80.50} &
          \multirow{2}{*}{76.16} &
          \multirow{2}{*}{71.78} &
          \multirow{2}{*}{86.65} &
          \multirow{2}{*}{78.79} \\
         &
          K={[}4, 4, 1, 1, 1, 1{]} &
           &
           &
           &
           &
           &
           &
           \\ \hline
        \end{tabular}
    }
    \label{tab: WeightLocking}
\end{table}

\begin{table*}[htp]
\caption{\small Main comparison results among our method with prior defense methods against BFA attackers.}
\vspace{-10pt}
\resizebox{0.92\textwidth}{!}{
    \begin{tabular}{c|c|c|c|c|ccc|cccc}
    \hline
     &
       &
       &
       &
       &
       &
       &
       &
      \multicolumn{1}{c|}{} &
      \multicolumn{3}{c}{Memory Overhead} \\
    \multirow{-2}{*}{\begin{tabular}[c]{@{}c@{}}Model + \\ dataset\end{tabular}} &
      \multirow{-2}{*}{Category} &
      \multirow{-2}{*}{\begin{tabular}[c]{@{}c@{}}Quant. \\ Bit\end{tabular}} &
      \multirow{-2}{*}{\begin{tabular}[c]{@{}c@{}}Defense \\ Method\end{tabular}} &
      \multirow{-2}{*}{\begin{tabular}[c]{@{}c@{}}Prior-attack \\ Accuracy\end{tabular}} &
      \multirow{-2}{*}{Best Acc.} &
      \multirow{-2}{*}{Worst Acc.} &
      \multirow{-2}{*}{\textbf{Mean Acc.}} &
      \multicolumn{1}{c|}{\multirow{-2}{*}{\begin{tabular}[c]{@{}c@{}}Training/Searching \\ Runtime\end{tabular}}} &
      Pre ($m_{TCU}$) &
      Post ($m_L$) &
      Total \\ \hline
     &
       &
      4-bit &
      - &
      87.73 &
      82.24 &
      59.87 &
      75.85 &
      \multicolumn{4}{c}{} \\
     &
       &
      6-bit &
      - &
      88.00 &
      75.66 &
      33.58 &
      61.74 &
      \multicolumn{4}{c}{} \\
     &
      \multirow{-3}{*}{w/o Def} &
      8-bit &
      - &
      88.00 &
      53.62 &
      13.52 &
      32.89 &
      \multicolumn{4}{c}{\multirow{-3}{*}{-}} \\ \cline{2-12} 
     &
       &
      1-bit &
      BAT~\cite{RADAR_DATE21_Li} &
      87.09 &
      86.12 &
      \multicolumn{1}{c}{74.62} &
      \multicolumn{1}{c|}{80.39} &
      \multicolumn{1}{c|}{0.33 hrs} &
      \multicolumn{3}{c}{} \\
     &
       &
      4-bit &
      NAT~\cite{NP_DATE2020_Gu} &
      87.96 &
      83.32 &
      66.71 &
      77.06 &
      \multicolumn{1}{c|}{2.8 hrs} &
      \multicolumn{3}{c}{} \\
     &
       &
      6-bit &
      NAT~\cite{NP_DATE2020_Gu} &
      87.19 &
      77.68 &
      33.39 &
      64.78 &
      \multicolumn{1}{c|}{2.8 hrs} &
      \multicolumn{3}{c}{} \\
     &
      \multirow{-4}{*}{Training-based} &
      8-bit &
      NAT~\cite{NP_DATE2020_Gu} &
      85.91 &
      67.74 &
      26.14 &
      55.88 &
      \multicolumn{1}{c|}{2.8 hrs} &
      \multicolumn{3}{c}{\multirow{-4}{*}{-}} \\ \cline{2-12} 
     &
       &
       &
      Pruning~\cite{RADAR_DATE21_Li} &
      87.73 &
      80.88 &
      57.23 &
      70.68 &
      \multicolumn{1}{c|}{-} &
      \multicolumn{3}{c}{3.13\% (G=16)} \\
     &
       &
      \multirow{-2}{*}{4-bit} &
      \textbf{Ours} &
      87.73 &
      86.96 &
      83.08 &
      {\color[HTML]{0070C0} \textbf{84.74}} &
      \multicolumn{1}{c|}{0.03hrs + 0.33 hrs} &
      0.84\% &
      0.000\% &
      0.84\% \\ \cline{3-12} 
     &
       &
       &
      Pruning~\cite{RADAR_DATE21_Li} &
      88.00 &
      79.91 &
      40.74 &
      66.14 &
      \multicolumn{1}{c|}{-} &
      \multicolumn{3}{c}{4.17\% (G=8)} \\
     &
       &
      \multirow{-2}{*}{6-bit} &
      \textbf{Ours} &
      88.00 &
      86.90 &
      86.25 &
      {\color[HTML]{0070C0} \textbf{86.48}} &
      \multicolumn{1}{c|}{0.03hrs + 0.50 hrs} &
      0.93\% &
      1.11\% &
      2.04\% \\ \cline{3-12} 
     &
       &
       &
      Pruning~\cite{RADAR_DATE21_Li} &
      88.00 &
      70.11 &
      18.68 &
      48.59 &
      \multicolumn{1}{c|}{-} &
      \multicolumn{3}{c}{3.13\% (G=8)} \\
    \multirow{-13}{*}{\begin{tabular}[c]{@{}c@{}}VGG-8\\  + \\ CIFAR10\end{tabular}} &
      \multirow{-6}{*}{Training-free} &
      \multirow{-2}{*}{8-bit} &
      \textbf{Ours} &
      88.00 &
      87.21 &
      86.08 &
      {\color[HTML]{0070C0} \textbf{86.73}} &
      \multicolumn{1}{c|}{0.03hrs + 0.75 hrs} &
      1.07\% &
      1.29\% &
      2.36\% \\ \hline
     &
       &
      4-bit &
      - &
      61.28 &
      56.49 &
      49.06 &
      54.76 &
      \multicolumn{4}{c}{} \\
     &
       &
      6-bit &
      - &
      60.61 &
      41.45 &
      3.69 &
      22.89 &
      \multicolumn{4}{c}{} \\
     &
      \multirow{-3}{*}{w/o Def} &
      8-bit &
      - &
      60.22 &
      38.01 &
      2.19 &
      9.93 &
      \multicolumn{4}{c}{\multirow{-3}{*}{-}} \\ \cline{2-12} 
     &
       &
      1-bit &
      BAT~\cite{DefendBFA_CVPR2020_He} &
      60.03 &
      59.69 &
      \multicolumn{1}{c}{54.01} &
      \multicolumn{1}{c|}{56.84} &
      \multicolumn{1}{c|}{0.4 hrs} &
      \multicolumn{3}{c}{} \\
     &
       &
      4-bit &
      NAT~\cite{NP_DATE2020_Gu} &
      61.58 &
      58.27 &
      46.72 &
      51.81 &
      \multicolumn{1}{c|}{8.9 hrs} &
      \multicolumn{3}{c}{} \\
     &
       &
      6-bit &
      NAT~\cite{NP_DATE2020_Gu} &
      60.61 &
      39.24 &
      7.20 &
      22.89 &
      \multicolumn{1}{c|}{8.9 hrs} &
      \multicolumn{3}{c}{} \\
     &
      \multirow{-4}{*}{Training-based} &
      8-bit &
      NAT~\cite{NP_DATE2020_Gu} &
      60.22 &
      38.98 &
      2.19 &
      12.99 &
      \multicolumn{1}{c|}{8.9 hrs} &
      \multicolumn{3}{c}{\multirow{-4}{*}{-}} \\ \cline{2-12} 
     &
       &
       &
      Pruning~\cite{RADAR_DATE21_Li} &
      61.28 &
      57.24 &
      42.02 &
      47.68 &
      \multicolumn{1}{c|}{-} &
      \multicolumn{3}{c}{3.13\% (G=16)} \\
     &
       &
      \multirow{-2}{*}{4-bit} &
      \textbf{Ours} &
      61.28 &
      60.20 &
      53.78 &
      {\color[HTML]{0070C0} \textbf{57.40}} &
      \multicolumn{1}{c|}{0.04 hrs + 1.17 hrs} &
      1.12\% &
      1.05\% &
      2.17\% \\ \cline{3-12} 
     &
       &
       &
      Pruning~\cite{RADAR_DATE21_Li} &
      60.61 &
      54.96 &
      10.99 &
      44.76 &
      \multicolumn{1}{c|}{-} &
      \multicolumn{3}{c}{4.17\% (G=8)} \\
     &
       &
      \multirow{-2}{*}{6-bit} &
      \textbf{Ours} &
      60.61 &
      59.46 &
      58.21 &
      {\color[HTML]{0070C0} \textbf{58.88}} &
      \multicolumn{1}{c|}{0.04 hrs + 1.62 hrs} &
      1.06\% &
      1.20\% &
      2.26\% \\ \cline{3-12} 
     &
       &
       &
      Pruning~\cite{RADAR_DATE21_Li} &
      60.22 &
      53.62 &
      16.33 &
      39.72 &
      \multicolumn{1}{c|}{-} &
      \multicolumn{3}{c}{3.13\% (G=8)} \\
    \multirow{-13}{*}{\begin{tabular}[c]{@{}c@{}}ResNet-18 \\ + \\ CIFAR100\end{tabular}} &
      \multirow{-6}{*}{Training-free} &
      \multirow{-2}{*}{8-bit} &
      \textbf{Ours} &
      60.22 &
      58.82 &
      57.21 &
      {\color[HTML]{0070C0} \textbf{58.10}} &
      \multicolumn{1}{c|}{0.04 hrs + 3.01 hrs} &
      1.26\% &
      1.77\% &
      3.03\% \\ \hline
    \end{tabular}
}
\label{tab:MainResults}
\vspace{-10pt}
\end{table*}

\vspace{-3pt}
\subsubsection{Memory Overhead Assignment in Pre-Attack Unary Protection}
\label{sec:AblationSparsity}
We compare \emph{Top-Sensitive-Layer Assignment} to an \emph{Even Assignment} baseline, i.e., $m_U^l=m_U/L$, in Table~\ref{tab:QU-Even}.
With unary-coded pre-attack protection, the even assignment method consumes significantly higher memory overhead than our sensitivity-aware method when reaching the same level of post-attack accuracy.

\vspace{-5pt}
\subsubsection{Truncated Complementary Unary Protection}

Table~\ref{tab:CQU-res} compares the memory storage required by Unary Protection and Truncated Complementary Unary Protection. 
TCU can significantly reduce the memory overhead by 6-11$\times$ than the original unary encoding.

\vspace{-3pt}
\subsubsection{Post-Attack Weight Locking}
For weight locking, the accuracy drop threshold $\eta$ is the key parameter to balance resumed accuracy and memory overhead.
In Table~\ref{tab: WeightLocking}, we show the performance of the proposed Weight Locking on 8-bit VGG-8 and CIFAR-10. 
Within the layer-wise acceptable accuracy drop $\eta$, Weight Locking can provide significant post-attack accuracy recovery with only less than 2\% of extra memory overhead.
However, to achieve lower memory overhead, accuracy recovery will be largely compromised from 87\% to 70\%. Weight Locking actually employs the accuracy-storage trade-off.

We also compare our method with Weight Pruning proposed in~\cite{RADAR_DATE21_Li}. 
Fig.~\ref{fig:CompareLockingPruning} shows the effectiveness and memory overhead of protection by Weight Pruning under different detection group sizes $G$.
Weight Locking can achieve higher accuracy recovery with more than 10$\times$ reduction in memory consumption.

\subsubsection{Optimal Combination of TCU and Locking}
\begin{figure}
    \centering
    \vspace{-5pt}
    \subfloat[]{\includegraphics[width=0.42\columnwidth]{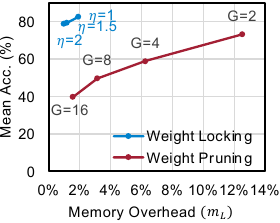}
    \label{fig:CompareLockingPruning}}
    \subfloat[]{\includegraphics[width=0.57\columnwidth]{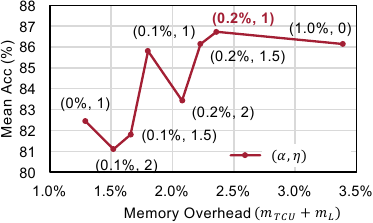}
    \label{fig:IntegratedSearching}}
    
    \vspace{-10pt}
    \caption{\small (a) Weight locking outperforms pruning~\cite{RADAR_DATE21_Li} with higher resumed accuracy and lower memory overhead.
    (b) Average resumed accuracy and memory overhead with various protection rates $\alpha$ in TCU and accuracy drop thresholds $\eta$ in locking with
    8-bit VGG8-CIFAR10.}
    \vspace{-10pt}
\end{figure}

Fig.~\ref{fig:IntegratedSearching} presents the searching process with different combinations of $(\alpha, \eta)$.
Pure unary protection will consume high memory overhead to achieve effective protection, while pure locking cannot offer comparable accuracy recovery.
$(\alpha,\eta )=(0.2\%,1)$ will give the optimal solution considering both the accuracy recovery and memory overhead.

\vspace{-5pt}
\subsection{Main Results}
\label{sec:MainResults}

In Table~\ref{tab:MainResults}, we compare our TCU+Locking scheme with NAT~\cite{NP_DATE2020_Gu} (std.=0.005 weight noise injection), BAT~\cite{DefendBFA_CVPR2020_He}, and pruning~\cite{RADAR_DATE21_Li}.
Our method can resume accuracy with only a 2\% drop after BFA attacks at a marginal 3\% memory overhead, significantly outperforming all prior arts.
Our method is training-free, which also saves significant runtime compared to training-based methods.

\vspace{-5pt}
\subsection{Discussion: Can Noises Become Defender?}
Besides low-bit quantization and sparsity, on-chip hardware noises are the main source of non-idealities.
While noises often degrade the inference accuracy, they also hinder the attack process by adding uncertainty to loss functions or gradients.
To reduce the uncertainty, attackers might need to average over multiple ($N_S$) samples, which equivalently reduces the attack efficiency.
\begin{table}
\centering
    \vspace{-0pt}
    \caption{\small BFA attack performance with different samples $N_S$ on 8-bit VGG8-CIFAR10. $BS$=16. }
    \vspace{-10pt}
    \resizebox{0.9\linewidth}{!}{
    \begin{tabular}{c|ccccc|c}
    \hline
        \multirow{2}{*}{Sample Times $N_S$} & \multicolumn{5}{c|}{Inference Budget $T_{inf}$}                                                & \multirow{2}{*}{Mean Acc} \\
          & 60 & 120 & 240 & 400 & 480 &        \\ \hline
        1                            & \textbf{44.94} & \textbf{34.83} & \textbf{20.88} & \textbf{16.14} & \textbf{13.61} & \textbf{26.08}           \\
        2 & 56.24   & 47.32   & 29.79   & 25.19   & 20.14  & 35.74 \\
        3 & 60.47   & 50.86   & 39.58   & 35.46   & 25.84  & 42.44 \\
        \hline
        \end{tabular}
    }
    \vspace{-10pt}
    \label{tab:noise-protection}
\end{table}

However, we find that sampling the noisy gradients (noise std.=0.005) once without averaging gives the best attack performance.

\begin{table}
\centering
\small
    \caption{\small Performance of adversarial attacker with different on-chip Gaussian weight noise level (std.) on 8-bit VGG8-CIFAR10.}
    \vspace{-10pt}
    \resizebox{\linewidth}{!}{
        \begin{tabular}{c|ccccccc}
        \hline
        noise std. $\backslash$ $T_{inf}$ & 0      & 40  & 180  & 400  & 600  & 800  & 900         \\ \hline
        0.005        & 87.58 & 52.45  & 35.18   & 18.59   & 17.02   & 15.67   & \textbf{13.52} \\
        0.01         & 86.82 & 59.68  & 34.64   & 18.42   & 16.15   & 31.88    & \textbf{21.77} \\
        0.02         & 83.81 & 54.87  & 32.12   & 21.97   & 17.21    & 34.19    & \textbf{14.48} \\
        0.03         & 74.69 & 41.06  & 25.82    & 20.74   & 17.29   & 20.45    & \textbf{18.10}   \\
        \hline
        \end{tabular}
    }
    \vspace{-10pt}
    \label{tab:HardwareNoise}
\end{table}

We increase the noise intensity in Table~\ref{tab:HardwareNoise} to create higher uncertainty for the BFA attacker.
Unfortunately, adding larger Gaussian noises to the weights during on-chip computations does not provide clear protection effects but leads to severe accuracy drops.
It is worth investigating potential protective noise injection and the trade-offs between noise-induced error and protection effects in the future.

\vspace{-5pt}
\section{Conclusion}
\label{sec:Conclusion}
In this work, \emph{for the first time}, we investigate the security issue of analog optical neural networks and present a novel nonideality-enabled built-in defender against adversarial bit-flip attacks.
We introduce quantization-inspired pre-attack protection based on truncated complementary unary weight representation to minimize the weight sensitivity with optimized memory overhead.
A complementary pruning-inspired weight-locking method is introduced to resume accuracy with precise error correction.
Our method outperforms prior defense approaches with near-ideal accuracy recovery under bit-flip attacks with marginal ($<$3\%) memory overhead.
Our work makes significant strides toward reliable ONN against adversarial weight attacks and unlocking future applications in security-thirst scenarios.



\end{document}